\documentclass[12pt]{iopart}  
\usepackage{algorithmic}
\usepackage{graphicx}
\usepackage{textcomp}
\usepackage[table,xcdraw]{xcolor} 
\usepackage{psfrag}
\usepackage{pdfpages}
\usepackage{subcaption}
\usepackage{array} 
\usepackage{multirow} 

\def\BibTeX{{\rm B\kern-.05em{\sc i\kern-.025em b}\kern-.08em
    T\kern-.1667em\lower.7ex\hbox{E}\kern-.125emX}}
    
\begin{document}
\newcommand{\sizeforsingleimages}{0.75} 

\title{Towards a quantum realization of the ampere using single-electron resolution Skipper-CCDs}

\captionsetup[subfigure]{labelformat=simple}
\renewcommand\thesubfigure{(\alph{subfigure})}

\author{Miqueas E. Gamero$^{1,2, 4}$, Agustin J. Lapi$^{1,2, 4}$, Blas J. Irigoyen Gimenez$^{1,2, 4}$, Fernando Chierchie$^{2,4}$, Guillermo Fernández Moroni$^{1}$, Brenda A. Cervantes-Vergara$^1$, Javier Tiffenberg$^1$, Juan Estrada$^1$, Eduardo E. Paolini$^{2, 3, 4}$, Gustavo Cancelo$^{1}$}

\address{$^1$ Fermi National Accelerator Laboratory, P.O. Box 500, Batavia, IL 60510, USA}
\address{$^2$ Instituto de Investigaciones en Ingeniería Eléctrica (IIIE) Alfredo Desages (UNS-CONICET), Departamento de Ingeniería Eléctrica y de Computadoras, Universidad Nacional del Sur (UNS), Bahía Blanca 8000, Argentina.}
\address{$^3$ Comisión de Investigaciones Científicas de la Provincia de Buenos Aires (CICpBA), La Plata 1900, Argentina}
\address{$^4$ Universidad Nacional del Sur (UNS), Bahía Blanca, Argentina}
\ead{miqueas.gamero@uns.edu.ar}

\begin{abstract}
This paper presents a proof-of-concept demonstration of the Skipper-CCD, a sensor with single-electron counting capability, as a promising technology for implementing an electron-pump-based current source. Relying on its single-electron resolution and built-in charge sensing, it allows self-calibration of the charge packets. This article presents an initial discussion of how low ppm and high current realizations can be achieved with this technology. We report experimental results that illustrate the key functionalities in manipulating both small and large electron charge packets, including a comparison of the charge generated, self-measured, and drained by the sensor against measurements from an electrometer. These results were obtained using a standard sensor and readout electronics without specific optimizations for this application. The objective is to explore the potential of Skipper-CCD for realizing an electron-based current source.
\end{abstract}

\section{Introduction}

With the recent redefinition of the unit of electric current, one ampere corresponds to the flow of $(1.602176634 \times 10^{-19})^{-1}$ elementary charges (electrons) per second, based on the fundamental defining constant of the elementary charge, $e^-=1.602176634 \times 10^{-19}$ C \cite{BIMP2019_Brochure}. This definition implies that a practical realization of the unit should control the number of electrons that are pumped or drained by a device within a given unit of time, and replaces the previous definition based on ampere's force law.

Over the past 35 years, significant efforts in the field of metrology have focused on single-electron current sources, making possible the generation of currents $I=f e^-$ by pumping single-electrons at a specified frequency $f$. A variety of realization techniques, along with some of the difficulties they impose, have been reviewed in \cite{Pekola2013Review}. 

Metallic tunnel junctions for single-electron pumps and semiconducting quantum dots are some of the common technologies explored because these methods leverage single-electron tunneling effects to control the transport of electrons. The first metallic single-electron devices were presented in the early 1990s. Later, more sophisticated devices, such as the hybrid turnstile based on single-electron transistors (SET) were tested \cite{Pekola2008,PhysRevLett.101.066801}. Quantum dots induced by surface acoustic waves were also explored in \cite{Shilton_1996}. Promising results have been reported in the above-mentioned studies. In some scenarios, currents of up to a few tens of picoamperes were observed. In recent years, currents of hundreds of picoamperes have been reported in \cite{giblin2012towards,giblin2020realisation} and even reached the nanoampere regime operating at several gigahertz \cite{yamahata2017high}.

Another approach applies Ohm's law to a quantum circuit, combining the quantum Hall resistance with the Josephson voltage to create a realization of the ampere \cite{BrunPicardPractical2016}. Also, other technologies are being proposed, for example a proof-of-concept using nanoscale MOSFETs in \cite{cheung2020nanoscale,Cheung2023FET} could potentially implement a room-temperature quantum current source. Another category is the self-referenced devices in \cite{Fricke2014Self}, which include built-in single-electron detectors to measure the charge that is being pumped.

A practical realization of the ampere requires increasing the levels of the generated current above $1$ nA. If a single electron-pump is used, $I=fe^-$, then very high frequencies are needed, in the order of several gigahertz. However, there is a practical limit to the operational frequency at which accurate single-electron transfer can be maintained. Therefore, to increase the current, a more feasible approach could be to enlarge the charge packet $I=N f e^-$ using multiple single-electron sources in parallel \cite{norimoto2024parallelization,Nakamura2024Universality}. 

This approach can allow reaching the required levels of current, but requires careful design, tuning, and matching of the devices. Another approach could be to use a device that can directly handle larger packets of electrons, $N e^-$, ideally from a single carrier to hundreds or even thousands of them, while operating at a lower frequency. A technology such as the Charge Coupled Device (CCD) meets the requirements stated above.

The Skipper-CCD is a special type of CCD with single electron counting capability. It has a floating gate in the sense node that can non-destructively measure the charge packet, reaching sub-electron levels of noise. The packet $N e^-$ is then drained to an output terminal at a given readout frequency. The charge packet is measured by the built-in amplifier of the sensor, and therefore the Skipper-CCD could potentially implement a self-reference current source. Moreover, it operates at a temperature on the order of $140$K, which can be considered a major advantage compared to other technologies that require a few Kelvin or sub-Kelvin operation. This reduced complexity of the system can enable the scalability of future systems, including not just one but several sensors in the same setup. Previous efforts in the field of particle physics have already made significant progress in the parallelization and operation of hundreds \cite{chierchie2023first} and toward thousands of this devices  \cite{Cervantes-Vergara_2023}. Furthermore, Skipper-CCDs and Skipper in CMOS sensors with multiple output stages are being developed \cite{botti2023fast,Lapi2024MAS16, Lapi2024SkpCMOS} and dedicated single devices with thousand of this output stages are feasible. This devices can precisely measure charge packets from single electron to hundreds or thousands of electrons in each amplifier and then drain the charge using a common terminal, therefore they could be considered as future candidate for realization of a current source.

In this paper we present a proof-of-concept towards a self-referenced quantum current source implemented with a CCD, in particular a Skipper-CCD. Using a standard sensor with only four amplifiers, we develop several different experiments to show the potential of this technology.

The organization of the paper is as follows. In Section \ref{sec:PropposedTechniqueAndPPM} we introduce the proposed approach and analyze different theoretical ppm scenarios that could be achieved with existent CCD technology. In Section \ref{subsec:absolute_calibration} we introduce the absolute self-calibration of the sensor in a wide charge range based on its single-electron resolution. In Section \ref{subsec:comparisson} an electrometer is connected to the drain of the Skipper-CCD to compare the self charge-measurement to that measured by an off-the-shelf instrument. Then, in Section \ref{subsec:nano-Ampere} the sensor is configured for faster readout and large charge packet mode in an experiment demonstrating the capability of reaching currents in the nanoampere range using a single sensor and also comparing to the electrometer measurements. Conversely, the experiment in Section \ref{sec:discreteCurrent} shows the Skipper-CCD performance in single-electron resolution mode, attaining discrete low levels of current with quantum steps, demonstrating its high dynamic range. A more ambitious experiment was conducted in Section \ref{sec:arbitraryCurrentGeneration} by using a flexible readout technique which allows setting of a different readout frequency for each charge packet. Although the frequency tuning capabilities are coarse in this implementation, we demonstrate that achieving arbitrary current waveforms is possible. Finally, Section \ref{sec:discussion} elaborates on the most important insights gained from the experiments conducted.

This series of experiments suggests the potential versatility and flexibility that this technology could offer. A future implementation that combines these characteristics using dedicated hardware for Skipper sensors with multiple amplifiers may lead to a practical realization of a current source based on the electron charge.

\section{Skipper-CCD and proposed methodology}
\label{sec:PropposedTechniqueAndPPM}

Skipper-CCDs can attain single-electron resolution by sampling the charge packet in one pixel multiple times \cite{Tiffenberg:2017aac}. This is possible because its readout architecture implements a floating gate in its output amplifier that measures the charge through capacitive coupling, which, unlike other CCDs, enables multiple non-destructive readout at the expense of longer readout times. As the sensor is read out sequentially, the charge is removed from the channel through a drain terminal generating an output current. The Skipper-CCDs have a high-dynamic range, from a single electron to thousands of electrons. The readout noise can be adjusted based on the number of samples taken per pixel, which is defined by the operator. 

\begin{figure}[h!]
    \centering
    \includegraphics[width=\sizeforsingleimages\linewidth]{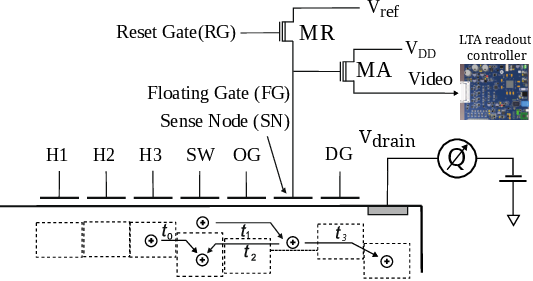}
    \caption{Schematic of the output stage of Skipper-CCD with its connections.}
    \label{fig:schematic-Skipper}
\end{figure}

A schematic of the output stage of a Skipper-CCD is shown in figure \ref{fig:schematic-Skipper}. A conventional readout sequence is shown alongside the schematic. The horizontal dashed lines depict the potential wells produced by different levels of voltages at the gates that occur during the sequence. These wells are proportional to the gate voltage. 

The first three gates, named H1, H2, and H3, are the three-phase clocks that move the charge through the horizontal register to the output amplifier. At instant $t_0$, the charge is transferred from H3 to the Summing Well (SW) gate. The SW and Output Gate (OG) facilitate the transfer of the charge packet to the Sense Node (SN), shown in time $t_1$. Then, the pixel charge value is measured from the video signal. The processing of the video signal and subsequent conversion to digital data is done in the Low Threshold Acquisition (LTA) readout controller \cite{cancelo2021low}. The charge can then be moved backward to enable a new independent measurement by repeating the process, as depicted at time $t_2$. The Reset Gate (RG) from the reset transistor (MR) is used to reset the SN level after each measurement. Finally, at instant $t_3$, the charge is drained by setting the Dump Gate (DG) voltage to low. An electrometer can be connected at this point so that the drained charge can be measured by two different techniques.

\begin{figure}[h!]   
    \psfrag{a}[c][c][0.7][0]{$T_0=1/f_0$}
    \psfrag{b}[c][c][0.7][0]{$T_1=N/f_0=1/f_1$}
    \psfrag{c}[l][c][0.7][0]{$t_{n_0}=1/f_{n_0}=n_0/(Nf_1)$}
    \psfrag{q}[c][c][0.7][0]{$q$}
    \psfrag{e}[c][c][0.7][0]{$e^{-}$}
    \psfrag{t}[c][c][0.7][0]{$t$}
    \psfrag{m}[c][c][0.7][0]{$t_{\mathrm{Med}}$}
    \psfrag{n}[c][c][0.7][0]{$Ne^{-}$}
    \psfrag{o}[c][c][0.7][0]{$n_0e^{-}$}
    \psfrag{p}[c][c][0.7][0]{$n_1e^{-}$}
    \psfrag{r}[c][c][0.7][0]{$n_2e^{-}$}
    \psfrag{s}[c][c][0.7][0]{$n_3e^{-}$}
    \psfrag{u}[c][c][0.7][0]{$n_4e^{-}$}
    \psfrag{v}[c][c][0.7][0]{$n_5e^{-}$}
    \psfrag{w}[c][c][0.7][0]{$n_ie^{-}$}
    \psfrag{x}[l][c][0.7][0]{$I=f_0e^{-}$}
    \psfrag{y}[l][c][0.7][0]{$I=Nf_1e^{-}$}
    \psfrag{z}[l][c][0.7][0]{$I=n_if_{n_i}e^{-}$}
    \centering
    \includegraphics[width=\sizeforsingleimages\textwidth]{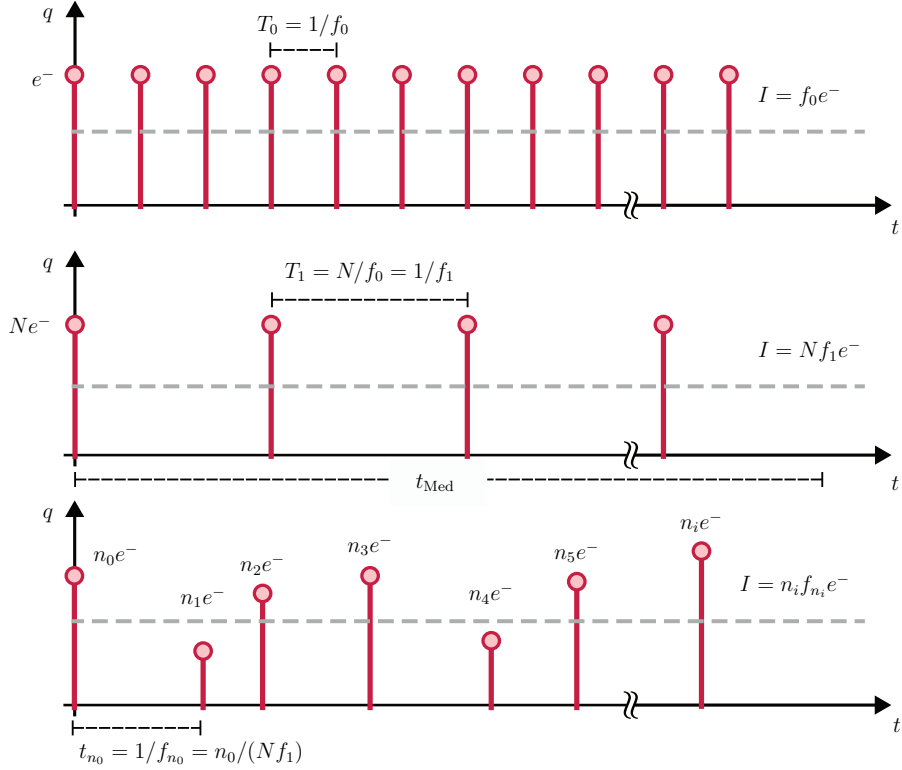}
    \caption{Different approaches for achieving the same average current. }
    \label{fig:impulsos}    
\end{figure}

In figure \ref{fig:impulsos}, the charge impulses depict possible sequences of different charge packets. In the case of the upper axis, each impulse represents a single electron, with a period or pixel acquisition time of $T_0$, resulting in a constant current of $I = e^-/T_0$. Throughout the article, the pixel rate is defined as the inverse of the period. In the middle axis, the same constant current is achieved by increasing the charge packet to $Ne^-$ electrons and by increasing the pixel period to $T_1 = N T_0$. The lower axis shows a third scenario, in which the pixel rate $t_{n_i}$ is also adjusted based on the charge packet size $n_ie^-$.

These three scenarios shows possible implementations of the same average quantum current. The first two are the more standard approaches while the last one is proposed as a realization that could potentially accommodate for different electron packets sizes $n_ie^{-}$ around a given mean $Ne^-$. This last approach requires to have a smart self-referenced system that can read the charge packet value and at the same time control the speed at which the packet is drained to generate the quantum current. The Skipper-CCD has all these capabilities, and this paper experimentally explores each of them separately.

\subsection{Theoretical achievable currents and precisions for different implementations}
\label{sec:ppm}

\begin{figure}[h!]
    \centering
    \includegraphics[width = \sizeforsingleimages\textwidth]{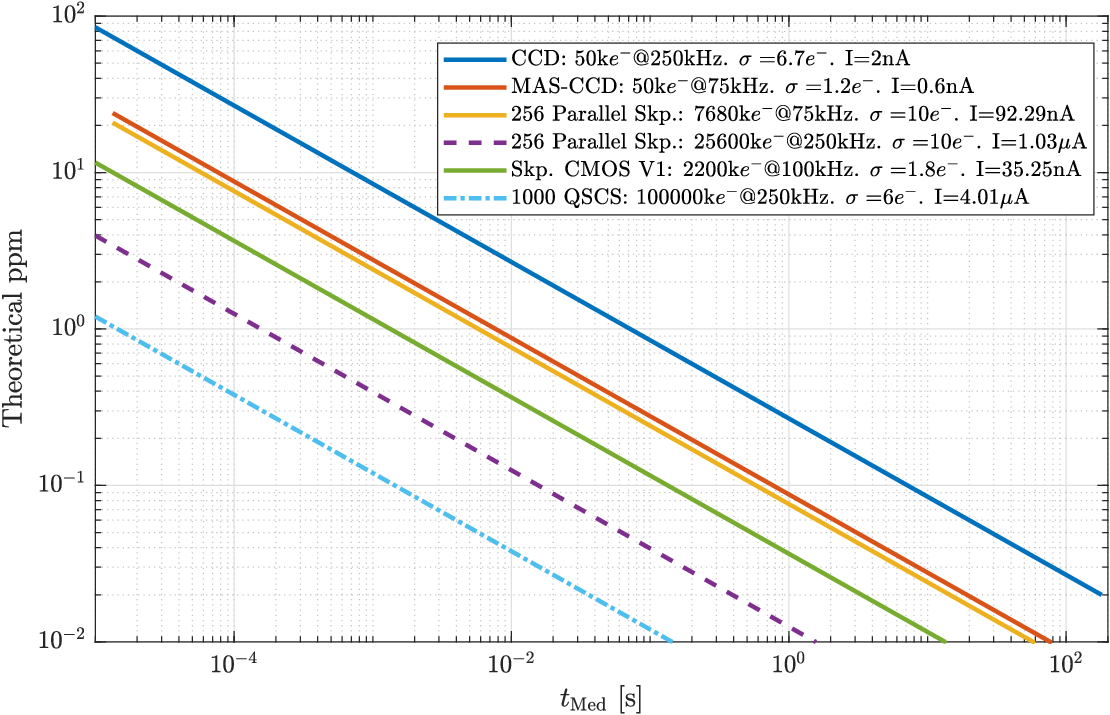}
    \caption{Theoretical ppm achievable with different sensors and possible experiments.}
    \label{fig:theoreticalPPM}
\end{figure}

Figure \ref{fig:theoreticalPPM} presents the theoretically achievable ppm values as a function of  the measurement time $t_{\mathrm{Med}}$ (as indicated in figure~\ref{fig:impulsos})  for various possible sensor operating conditions. The experimental conditions used to generate the simulation were extracted from previously published results in the literature related to CCDs and Skipper sensors, the ppm are calculated based on the uncertainty in measuring the charge packets and assume a perfect timing with period $T_1$. Additionally, two hypothetical scenarios indicates with dashed lines are presented.

In \cite{Lapi2023} standard CCD sensors are operated at pixel rate of $250$ kpix/s, resulting in $T_1=4$ $\mu$s with a noise of $\sigma=6.7$ $e^{-}$, assuming a charge packet of $N=50$ k$e^{-}$ and a $t_{\mathrm{Med}}=M T_1$ the ppm achieved after $t_{\mathrm{Med}}$ can be computed as
\setlength{\mathindent}{130pt}
\begin{equation}
\mathrm{ppm}=10^6\times \frac{\sqrt{M}\sigma}{N M}=10^6\times \frac{\sigma}{N \sqrt{M}}.
\end{equation}
During $t_{\mathrm{Med}}$, $M$ packages of $N$ electrons of charge are accumulated (denominator) while the noise after the addition of the charge packages scales with $\sqrt{M}$ (numerator). The blue curve identified as CCD in figure~\ref{fig:theoreticalPPM} shows the theoretical ppm for sensor performance in \cite{Lapi2023}, ppms follow a straight line in the log-log plot due to resulting $1/\sqrt{M}$ factor of reduction.

In another scenario, the Multiple-Amplifier Sensing CCD (MAS-CCD) which has $16$ inline Skipper amplifiers, can be used. The MAS-CCD has the ability to reduce the noise by measuring the same charge packet with the multiple inline amplifiers, without increasing the readout time per pixel. In \cite{Lapi2024MAS16} $T_1=13.3$ $\mu$s with a noise of $\sigma=1.2$ $e^{-}$ resulting in the red curve shown in figure~\ref{fig:theoreticalPPM}. In this scenario $0.01$ ppm can be achieved in $\sim 70$ seconds of measurement with a single sensor.

An experiment with parallel operation of $P$ sensors is described in \cite{chierchie2023first}. This instrument can operate and readout $P=256$ Skipper-CCDs. With this system, all the charge packets generated by each of the sensors are added at the drain terminal. For example if $30$ k$e^{-}$ are generated in each sensor with conservative  $T_1=40$ $\mu$s and an individual noise of each sensor of $\sigma=10$ $e^{-}$ for only one Skipper sample per pixel, then, when the charge packets of the $P$ sensors are added together the noise uncertainty of the $P \times N$ electrons charge packet is $\hat{\sigma}=\sqrt{P}\sigma$. Therefore the ppm can be computed as
\setlength{\mathindent}{160pt} 
\begin{equation}
\mathrm{ppm}=10^6\times \frac{\sqrt{MP}\sigma}{N M P}.
\end{equation}
In this scenario a current of $92.29$ nA can be achieved by reaching $0.01$ ppm in approximately one minute of measurement.
If the same system is operated at a pixel rate of $1/T_1=250$ kpix/s with $100$ k$e^{-}$ and the same readout noise, then in this hypothetical scenario, a current of $1$ $\mu$A is achievable (indicated with a dashed line in figure~\ref{fig:theoreticalPPM}).

The theoretical ppm for a Skipper imager in CMOS as the one recently reported in \cite{Lapi2024SkpCMOS} is also shown. For this ASIC each pixel has a Skipper output stage capable of sub-electron readout noise, hence increasing the readout speed and achievable current with a single device. Considering a conservative example where 200 pixels are read and the charge is drained simultaneously with a full well of $11000$ $e^-$ the green curve is obtained with a single sensor. This example has the potential of scaling and even room-temperature operation in future implementation and experimental validations.

Finally an hypothetical device containing $1000$ Skipper stages in parallel is simulated, indicated as Quantum Skipper Current Source (QSCS). Assuming this device can handle $100$ k$e^{-}$ per output stage at a readout frequency of  $1/T_1=250$ kpix/s, with an individual noise of $6$ $e^{-}$ per output stage, then a current of $4$ $\mu$A is achieved with $0.01$ ppm in less than one second of measurements.

In the following section, we use an standard Skipper-CCD sensor with four readout channels to experimentally demonstrate some of the key points previously described, moving toward a future realization of a quantum current source using this technology. 

\section{Methods and results}
\label{sec:ExperimentalResults}

\begin{figure*}[h!]
    \centering
    \includegraphics[width=0.95\textwidth]{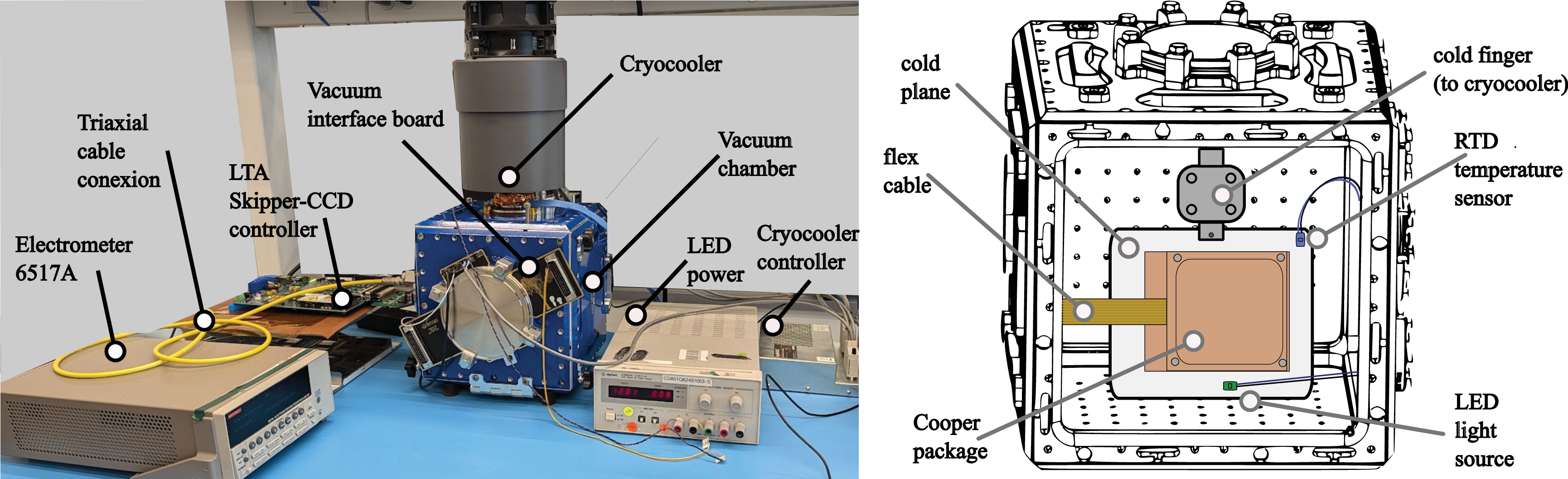}
    \caption{Experimental setup}
    \label{fig:setupExperimental}
\end{figure*}

Figure \ref{fig:setupExperimental} shows a picture of the experimental setup on the left and a drawing of the main components inside the Dewar on the right. The system is composed of a blue vacuum cube operating at $\sim 10^{-4}$ Torr. A cryocooler mounted on top of the cube maintains the sensor at $\sim 140$K. The cube has several vacuum interface electrical ports used to connect the LTA controller and other electrical components such as the RTD temperature sensors and the LED light source used for the experiments. The Skipper-CCD is a fully-depleted sensor developed and designed by Lawrence Berkeley National Laboratory (LBNL). It is a $\sim 700$ $\mu$m thick sensor with 1.35 megapixels, and is readout using the four available amplifiers, one in each corner.

The drawing on the right of figure~\ref{fig:setupExperimental} shows the main components inside the Dewar. The sensor is housed in a copper tray which is attached to a cold plane connected to the cold finger of the cryocooler. Inside the Dewar, there is also an RTD sensor and a controlled LED light source.

The Skipper-CCD is connected via a flex cable to a small passive PCB inside the Dewar, which then connects to a 50-pin vacuum feedthrough port. On the outside, there are two boards: one allows measurement of the charge drained from the sensor using a triaxial connection to the Keithley 6517A Electrometer, and another PCB with four pre-amplifiers for the CCD’s video signals, which then connects to the LTA controller that generates all the necessary bias and clock sequences to read the Skipper-CCD. None of the PCBs, flex cables, or the controller were optimized for this application. Additionally, the drain bias voltage, which is connected to the gate in the sensor where the charge is drained, is generated in the LTA and passes through all the aforementioned circuits; therefore, some leakage current is expected.

\subsection{Absolute calibration of the sensor for self-charge measurements}
\label{subsec:absolute_calibration}

Precise calibration is required to relate the analog to digital converter units (ADU) obtained at the readout controller to the charge packets in $e^-$. The Skipper-CCD, being a self-referenced system, can calibrate itself without needing an external calibration source. Its ability to resolve single electrons across a wide dynamic range enables precise calibration for each electron count, $n_ie^-$.

\begin{figure}[h!]
    \centering
    \includegraphics[width = \sizeforsingleimages\textwidth]{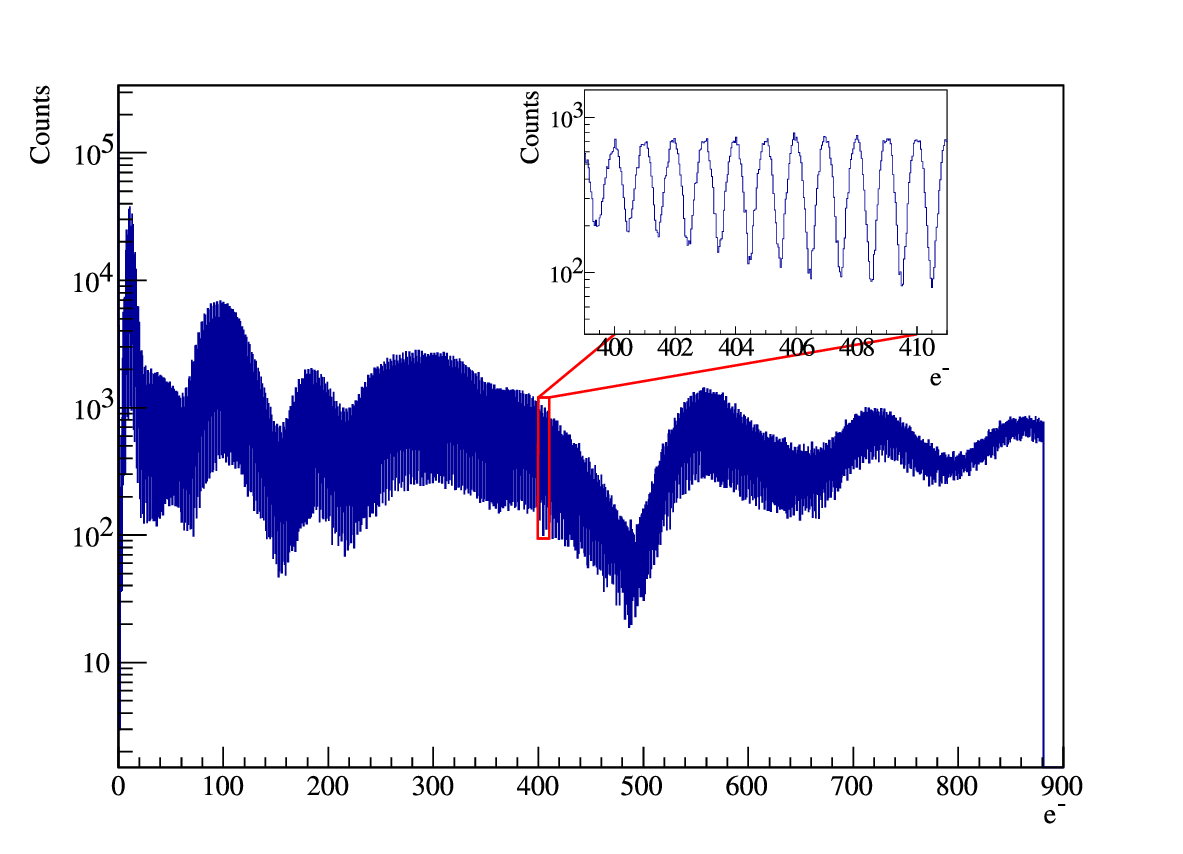}
    \caption{Histogram used for absolute calibration of Skipper-CCD.}
    \label{fig:histogram_e}
\end{figure}

The sensor operating in single-electron resolution with $\mathrm{NSAMP} = 1000$ samples per pixel was exposed to $16$ different light intervals ranging between $0$ and $30$ seconds. The $16$ resulting digital images, each corresponding to a different exposure, were combined into a single histogram that is depicted in figure \ref{fig:histogram_e}. Since photon arrivals follow Poisson statistics, each exposure produces an independent distribution with a characteristic shape that appears multiple times across the charge range.

In the zoom-in view, a range of eleven electron peaks between $400$ $e^-$ and $410$ $e^-$ is shown, with each electron count clearly distinguishable. Absolute calibration is achieved by fitting each Gaussian distribution, starting with the first Gaussian on the left, which represents $0$ $e^-$ measurements. An algorithm fits and counts each electron peak, creating a look-up table that relates ADUs to the electron charge packet. In this case, the calibration includes up to $910$ $e^-$. This calibration is used in the subsequent experiments to determine the charge measured by the Skipper-CCD, unless otherwise specified.

\subsection{Comparison of charge generated and self-measured by Skipper-CCD with electrometer measurements}
\label{subsec:comparisson}

\begin{figure*}[h!]
    \centering
    \includegraphics[width=\sizeforsingleimages\textwidth]{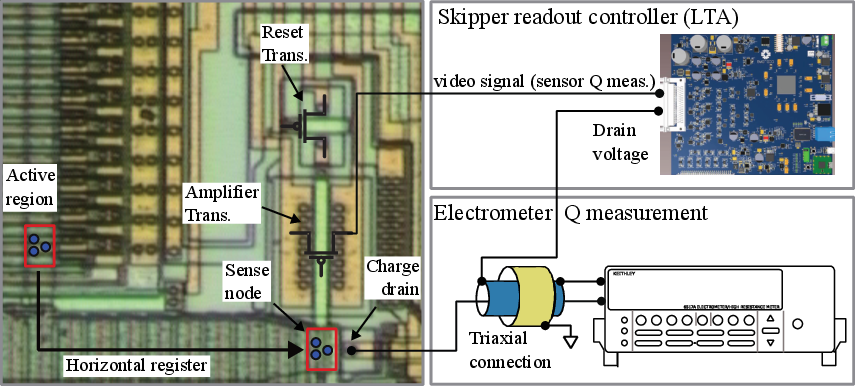}
    \caption{Schematic diagram and sensor micro photograph indicating the main components.}
    \label{fig:sensorPicAndSchematic}
\end{figure*}

The objective of this experiment is to demonstrate that the charge pumped out from the Skipper-CCD is compatible with the charge measured with a precision electrometer. A schematic showing the connections is presented in figure \ref{fig:sensorPicAndSchematic}. On the left, the image depicts a micro-photograph of the output stage of the Skipper-CCD, including a small portion of the active region and the horizontal register that ends at the sense node. As mentioned, the charge packet can be read non-destructively, either once or multiple times, using the built-in amplifier transistor that is capacitively coupled to the sense node. The sense node is then reset using the reset transistor, and the charge is finally drained out of the sensor. The entire readout sequence is controlled by the LTA controller, which also generates the drain gate voltage. This drain terminal is connected through a Keithley 6517A electrometer using a triaxial connector to measure the charges drained from the sense node. Thus, in this configuration, both the CCD and electrometer measure the generated charge, enabling a comparison between the performance of the two devices. 

\subsubsection{Methodology}

The electrometer is set to a range of $200$ nC, reset, and then starts recording at $\sim 2$ samples per second. Immediately after the reset, the LED inside the dewar is turned on during the desired exposure time. Once the exposure is complete, the Skipper-CCD readout begins. The readout was performed at a pixel rate of $\sim 26 \ \mathrm{kpix/s}$. A binning of $10$ rows, which implies adding $10$ rows of the active region into the horizontal register before reading to achieve higher charge levels per pixel, was applied \cite{janesick2001scientific}. The electrometer continues to acquire data even after the sensor readout is complete, during data conversion in the LTA controller. This experiment is repeated with light exposures of $0$, $3.75$, $7.5$, $11.25$, $15$, and $30$ seconds, generating increasing levels of charge proportional to the exposure time.

The raw electrometer data exhibits a consistent leakage current of approximately $10$ pA, which may be attributed to the printed circuit boards of the electronics involved in the setup. Furthermore, the drain voltage is generated by the LTA readout controller, located outside the Dewar and far from the sensor. The drain gate is polarized to $-22$ V, thus resulting in an equivalent leakage resistance to ground of $2200$ G$\Omega$. None of these subsystems has yet been optimized for low leakage current; future experiments will include the generation of the drain voltage next to the sensor in the same package and careful design of layouts and PCB materials.

The electrometer data is processed by fitting an interval where no charge is drained to measure leakage current, then subtracting this fit from each capture to remove the leakage effect. The total charge accumulated during each instance of the experiment was computed by subtracting the final charge value from the initial one.

As described in Section \ref{subsec:absolute_calibration}, an absolute calibration of the sensor was conducted. To convert the CCD electrons data into charge units, the conversion factor $1e^-=1.60217646\times10^{-19}$ C was used. Finally, the total charge accumulated from the four amplifiers in each corner of the Skipper-CCD was combined to enable comparison with the electrometer. 

\begin{figure*}[h!]
    \centering
    \includegraphics[width=\sizeforsingleimages\textwidth]{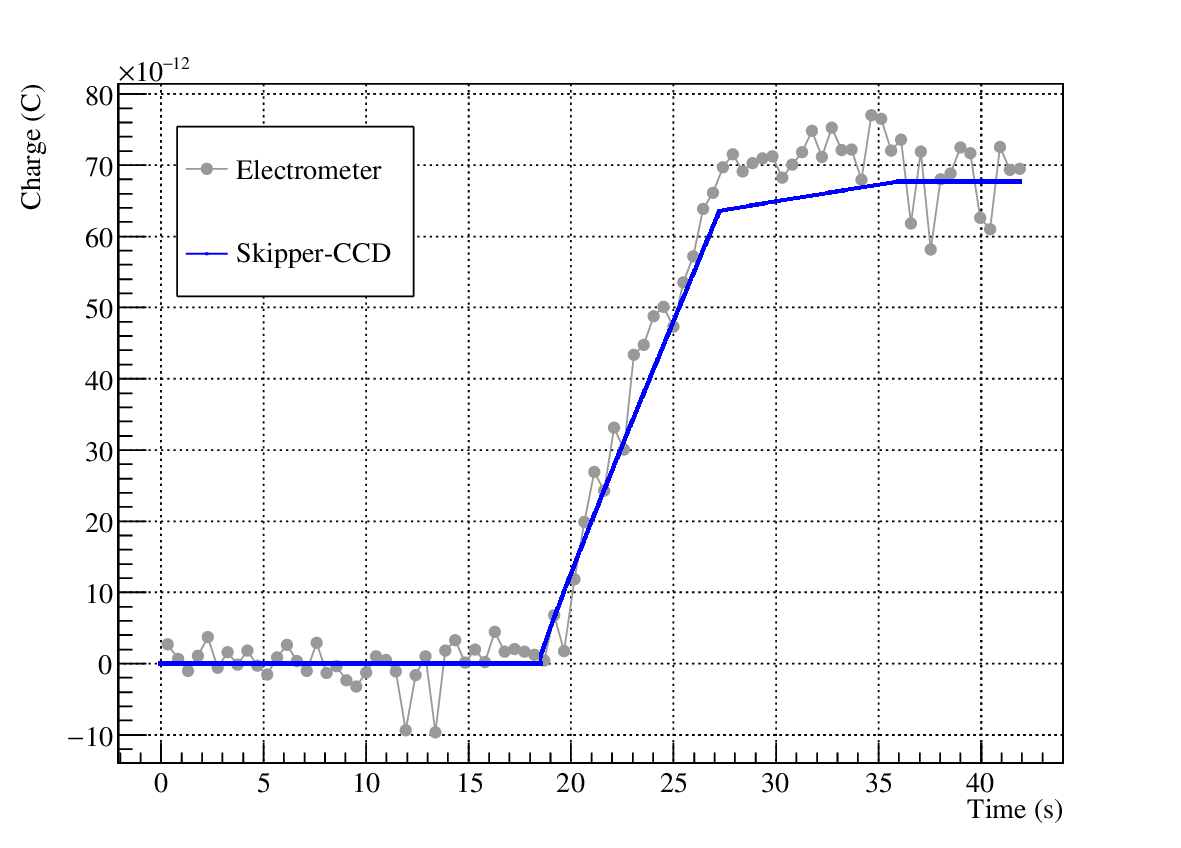}
    \caption{Charge measured by the Skipper-CCD and electrometer for an exposure time of $15$ s.}
    \label{fig:timePlot}
\end{figure*}

\subsubsection{Results}
\label{comparisson_results}

The electrometer and Skipper-CCD data, after processing, for a single dataset for an exposition of $15$ seconds are plotted together in figure \ref{fig:timePlot}. The first flat time-interval reflects the time during which the sensor is exposed to light plus instrument initialization. Then, the steep slope corresponds to the readout of the active region of the sensor, whereas the change of slope in the CCD data represents an over-reading of the sensor that drains less charge. In the final plateau, the sensor has been fully read, and  the conversion of the data is taking place in the LTA controller.

\begin{figure}[h!]
    \centering
    \begin{subfigure}[b]{0.49\textwidth}
        \centering
        \includegraphics[width=\textwidth]{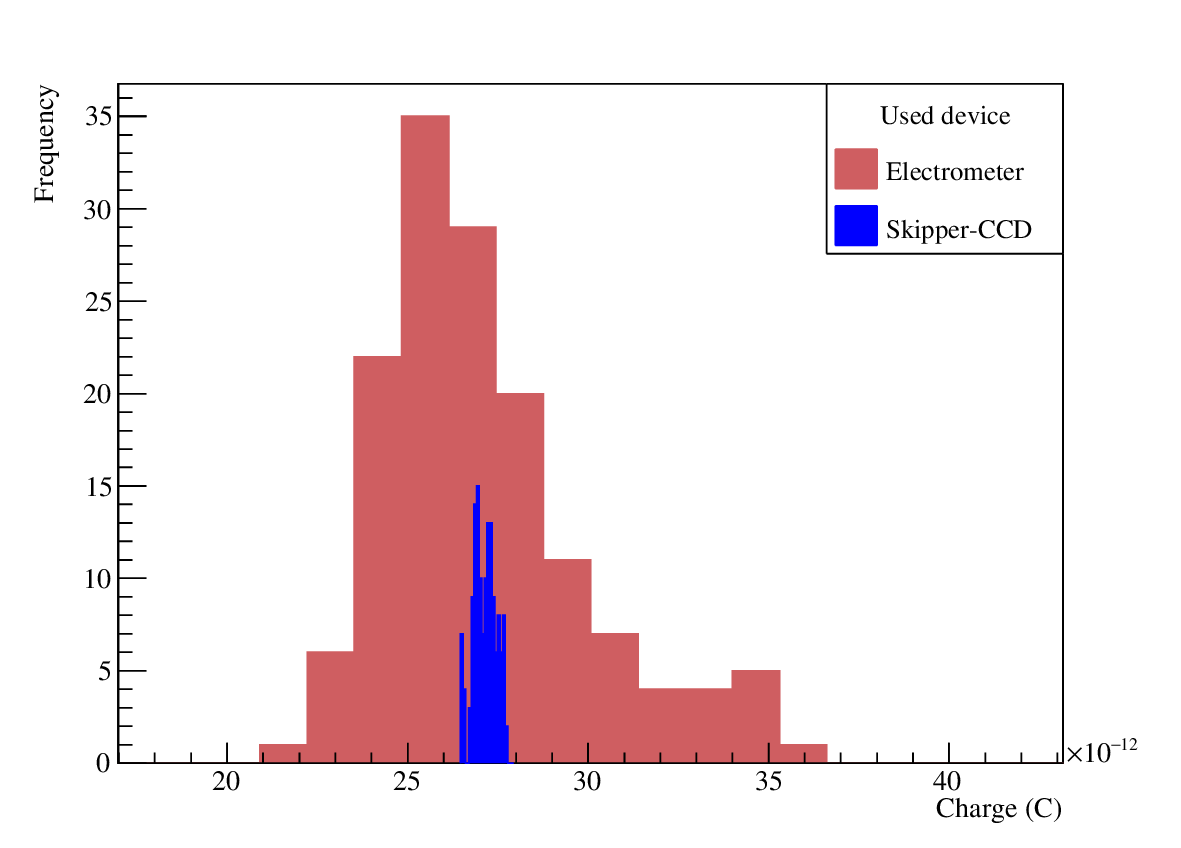}
        \caption{Histogram for charge using an electrometer and a Skipper-CCD.}
        \label{fig:charge_histogram}
    \end{subfigure}
    \hfill
    \begin{subfigure}[b]{0.49\textwidth}
        \centering
        \includegraphics[width=\textwidth]{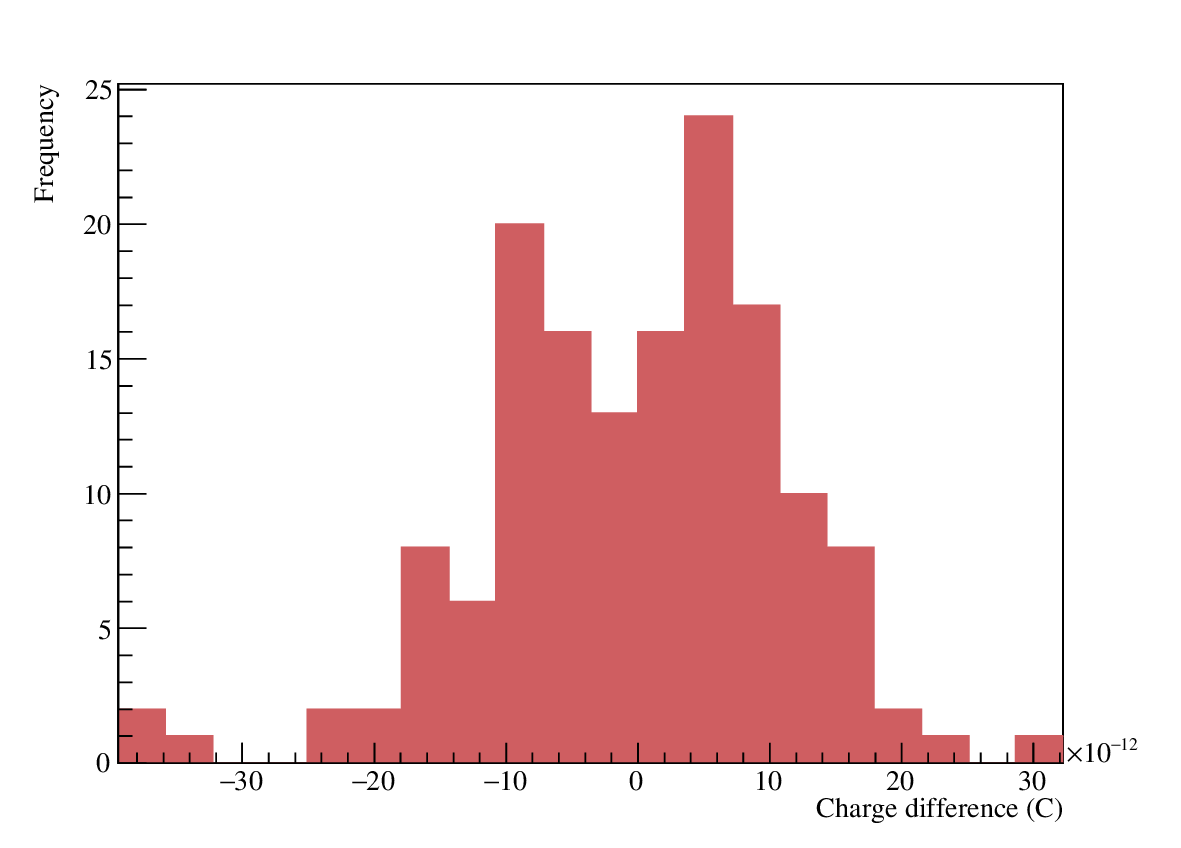}
        \caption{Histogram for the difference in charge measurements.}
        \label{fig:delta_charge_histogram}
    \end{subfigure}
    \caption{Comparison of charge measurements and their differences using an electrometer and a Skipper-CCD at exposure time of $3.75$ s.}
    \label{fig:histograms_electrometer_n_CCD}
\end{figure}

From $N_{\mathrm{cap}} = 144$ independent takes of the experiment described above, a set of histograms for each exposure time like those shown in figure \ref{fig:charge_histogram} were generated. Additionally, figure \ref{fig:delta_charge_histogram} shows a histogram after subtracting the CCD data from the electrometer data. The statistical results for each exposure are summarized in table \ref{tab:stat_results}, with the mean and standard deviation presented for both devices. The mean estimation introduces an error that was calculated as the square root of the variance of the mean, scaled by the square root of the number of samples: $e_\mu = \sigma/\sqrt{N_{\mathrm{cap}}}$. The statistics of the difference of the electrometer and Skipper-CCD measurements, along the associated estimation errors, are also included with the error propagation formula applied as $e_{\Delta,\mu} = \sqrt{\sigma_{\mathrm{elec}}^2/N_{\mathrm{cap}} - \sigma_{\mathrm{CCD}}^2/N_{\mathrm{cap}}}$. This expression is obtained under the assumption that no measuring nor processing error is associated to the CCD data. This premise will be further discussed in Section \ref{sec:discussion}.

\begin{table*}[h!]
    \centering
    \caption{Statistical results for different charge levels under different exposures}
    \renewcommand{\arraystretch}{1.3} 
    \setlength{\tabcolsep}{10pt} 
    \rowcolors{4}{white}{white} 
    \resizebox{\textwidth}{!}{ 
    \begin{tabular}{|c|c c|c c|c c c| c|}
        \hline
        \multirow{2}{*}{\textbf{Exposure [s]}} & \multicolumn{2}{c|}{\textbf{Charge at Electrometer [pC]}} & \multicolumn{2}{c|}{\textbf{Charge at CCD [pC]}} & \multicolumn{3}{c|}{\textbf{Difference [pC]}} \\
        \cline{2-8}
        & $\mu_{\mathrm{elec}}$ ($\pm e_\mu$) & $\sigma_{\mathrm{elec}}$ & $\mu_{\mathrm{CCD}}$ ($\pm e_\mu$) & $\sigma_{\mathrm{CCD}}$ & $\mu_{\mathrm{diff}}$ ($\pm e_{\Delta,\mu}$) & $\sigma_{\mathrm{diff}}$ & $\mu^{*}_{\mathrm{diff}}$ \\
        
        \hline
        0  & 0.35 (0.26) & 3.08 & -0.01 (0.00) & 0.00 & 0.12 (0.26) & 3.08 & 0.45 \\
        3.75  & 27.09 (0.24) & 2.87 & 27.13 (0.03) & 0.32 & -0.11 (0.24) & 2.80 & -0.74 \\
        7.5   & 42.56 (0.24) & 2.92 & 40.58 (0.05) & 0.54 & 1.94 (0.24) & 2.87 & 0.89 \\
        11.25 & 56.17 (0.26) & 3.02 & 53.89 (0.07) & 0.79 & 2.23 (0.25) & 2.92 & 0.85 \\
        15    & 69.89 (0.23) & 2.72 & 67.24 (0.08) & 0.99  & 2.59 (0.21) & 2.50 & 0.86 \\
        30    & 123.96 (0.21) & 2.56 & 120.79 (0.17) & 2.06 & 3.12 (0.13) & 1.52 & -0.01  \\
        \hline
    \end{tabular}
    }
    \label{tab:stat_results}
\end{table*}

\subsection{Achieving nanoampere currents using a single Skipper-CCD}
\label{subsec:nano-Ampere}

In this section we show that it is possible to achieve larger currents using a single Skipper-CCD. 
The connections for this experiment were the same as those shown in figure \ref{fig:sensorPicAndSchematic} and the acquisitions and processing are the same described in the previous Section \ref{subsec:comparisson}.

In this case, the electrometer was set to the same range of $200$ nC and sampling rate of $\sim 2$ samples per second, but the readout sequence of the Skipper-CCD was modified. To achieve currents up to the nanoampere range, the pixel rate was increased to $\sim 66.3 \ \mathrm{kpix/s}$, at the expense of higher readout noise, which was increased to a standard deviation of $40 \ e^-$, still very small in terms of these current levels. In addition, the binning of rows was duplicated to $20$ to yield higher currents by draining more charge per unit time. The experiment was repeated for light exposures of $20$, $40$, $60$, and $80$ seconds, generating increasing charge levels proportional to the exposure time.

For this large charge packet readout scheme, the Photon Transfer Curve (PTC) method was used for calibration \cite{janesick2001scientific}. For these results, the data is displayed in ampere units. This was done by carefully reconstructing the readout sequence. It is standard practice to over-read the CCD to create a `dark' region, known as the over-scan region, which is useful for measuring important parameters such as the readout noise or for subtracting the offset level from the images. During this over-reading stage, no charge is drained. Thus, the current estimation only considers the interval during which charge is being drained, which was calculated to be $t_{\mathrm{CCD}} = 287.30$ ms. Finally, the current is computed as $I = Q/t_{\mathrm{CCD}}$.

\begin{figure*}[h!]
    \centering
    \includegraphics[width=\sizeforsingleimages\textwidth]{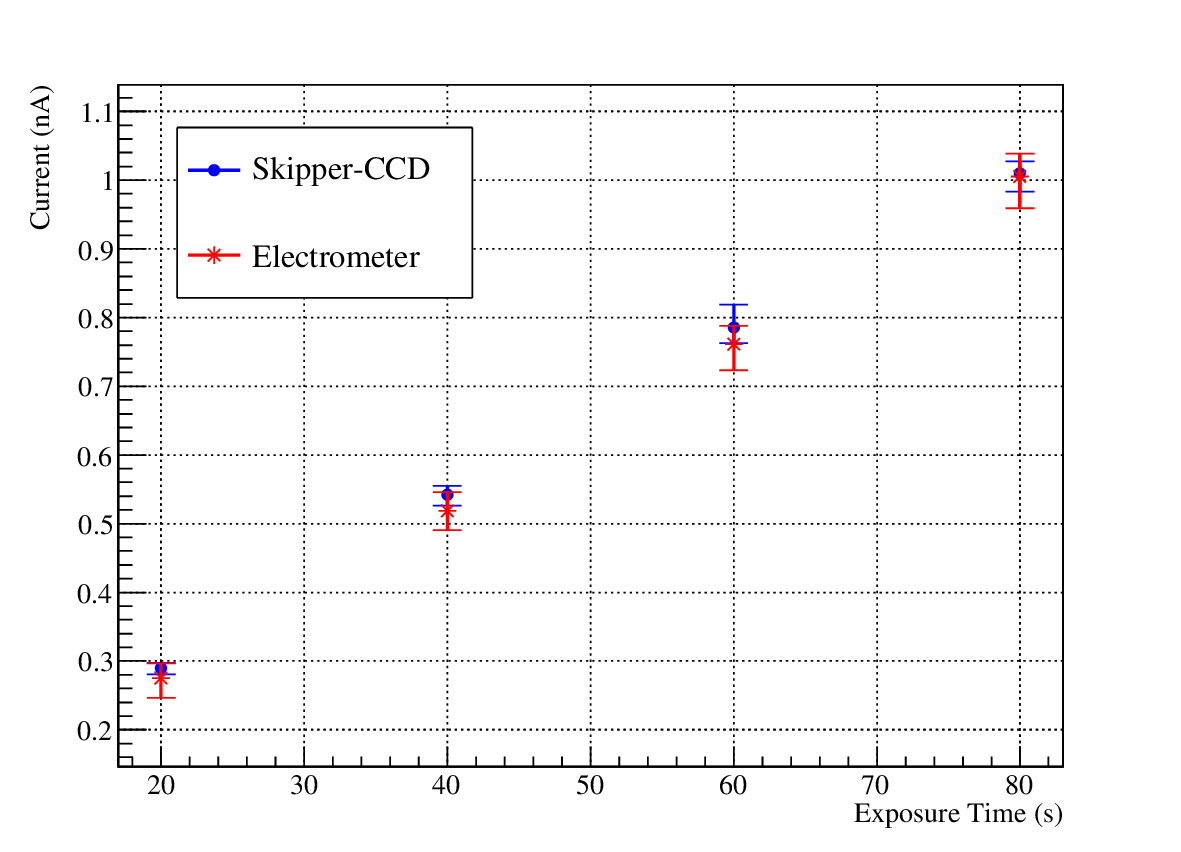}
    \caption{Larger currents achieved at different light exposure in fast-readout scheme.}
    \label{fig:nA_plot}
\end{figure*}

Figure \ref{fig:nA_plot} shows the current levels reached for each exposure. The experiment was repeated $150$ times for each exposure. The median of the data is represented by a circle for the CCD data and by an asterisk for the electrometer, and the minimum and maximum values are represented by vertical lines terminated in horizontal lines.

\subsection{Discrete current generation with quantum-steps}
\label{sec:discreteCurrent}

This experiment aims to demonstrate the capability of a Skipper-CCD to achieve discrete levels of current. The sensor was operated at a readout pixel rate of $\sim 67.83 \ \mathrm{pix/s}$ using a set of voltages that push the sensor into a condition where spurious charge is generated, normally undesired in imaging applications. To increase the charge per pixel, a row binning of $20000$ was employed. By taking $\mathrm{NSAMP} = 500$ samples per pixel, a readout noise of $\sim 0.18 \ e^-$ was attained, yielding single-electron resolution. The methodology is simple: the sensor is read without being exposed to light, as all the charge is generated spuriously in the sensor. In this case, a single take is enough to construct the dataset presented in this section.

To create a time series plot of the current, 300 consecutive pixels from the same row were considered. The pixel acquisition time, $T = 14.74$ ms, was used to reconstruct the time vector and to convert charge measurements into current units as $I = Q/T$. A histogram distinguishing individual discrete current steps was generated from a histogram of a region of the image. 

\begin{figure*}[h!]
    \centering
    \includegraphics[width=0.79\textwidth]{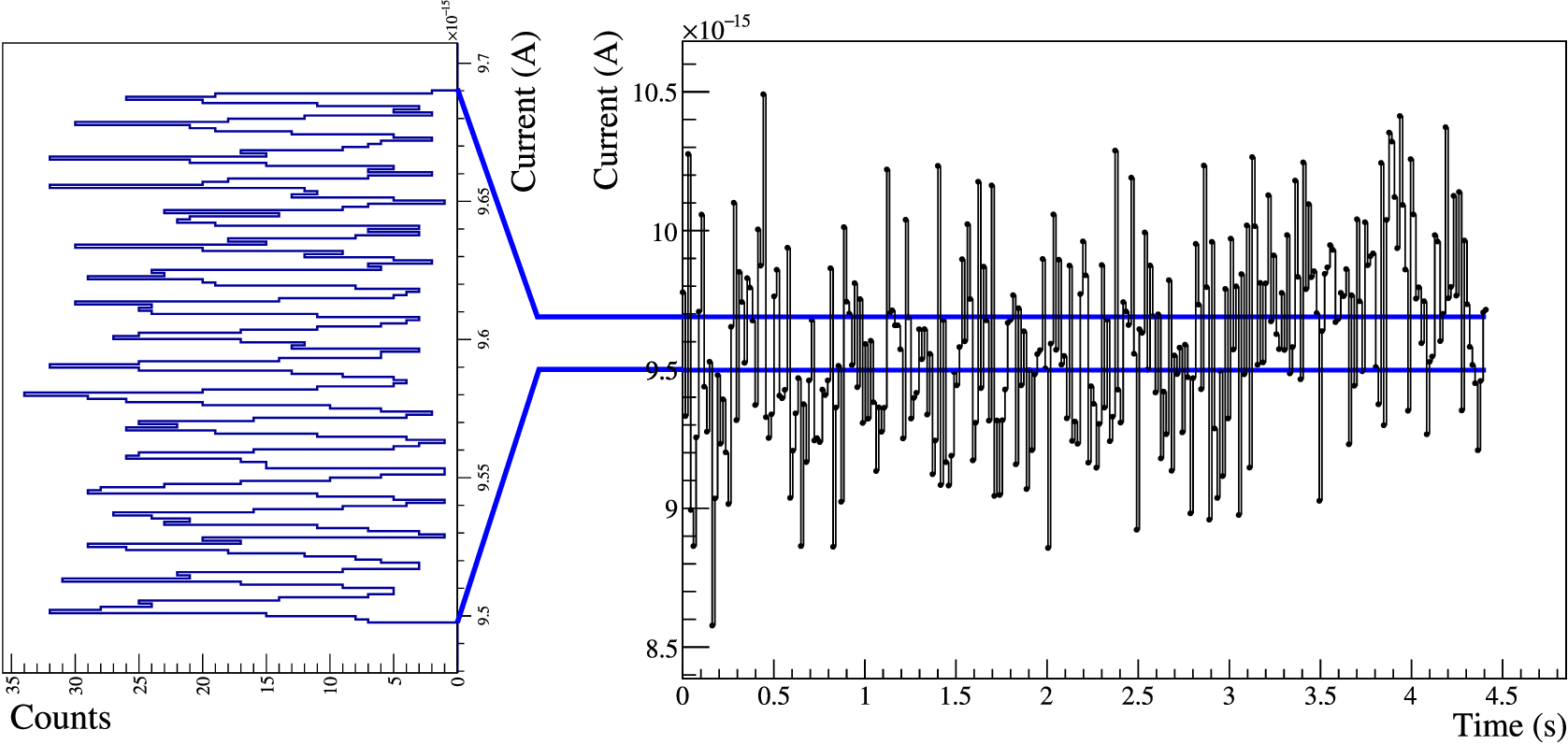}
    \caption{Discrete current levels achieved with Skipper-CCD.}
    \label{fig:discrete_current_peaks}
\end{figure*}

Figure \ref{fig:discrete_current_peaks} shows the results. To the right of the image, a zero-hold plot of current over time is depicted, showing a constant current level with a mean of $I_{\mathrm{mean}} = 9.59$ fA. The horizontal blue lines correspond to a range of $\pm 10\%$ around this mean. The histogram on the left shows the discretization of current within that range. This histogram was generated from a larger region of the image to increase the statistical reliability of the plot. 

\subsection{Arbitrary current generation}
\label{sec:arbitraryCurrentGeneration}

The Skipper-CCD and LTA controller can also be used to generate arbitrary current values by controlling when the charge packets are drained. As a proof-of-concept of this capability, we used the available tools of the Skipper-CCD readout system to generate arbitrary current waveforms by controlling the number of delay cycles used to dump the charge of each pixel into the drain gate. 

\begin{figure}[h!]
    \centering
    \begin{subfigure}[b]{0.49\textwidth}
        \centering
        \includegraphics[width=\textwidth]{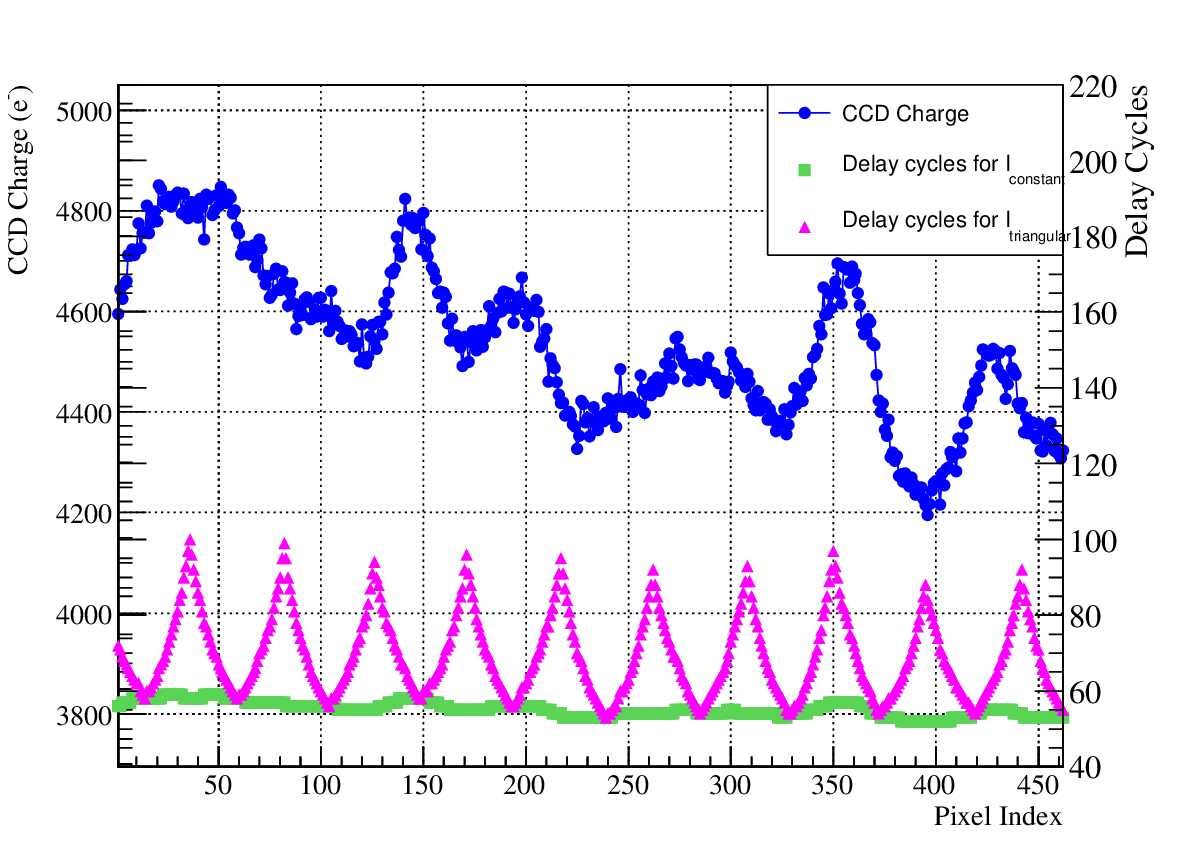}
        \caption{Charge measurements and off-line calculation of delays cycles.}
        \label{fig:charge_delaycicles}
    \end{subfigure}
    \hfill
    \begin{subfigure}[b]{0.49\textwidth}
        \centering
        \includegraphics[width=\textwidth]{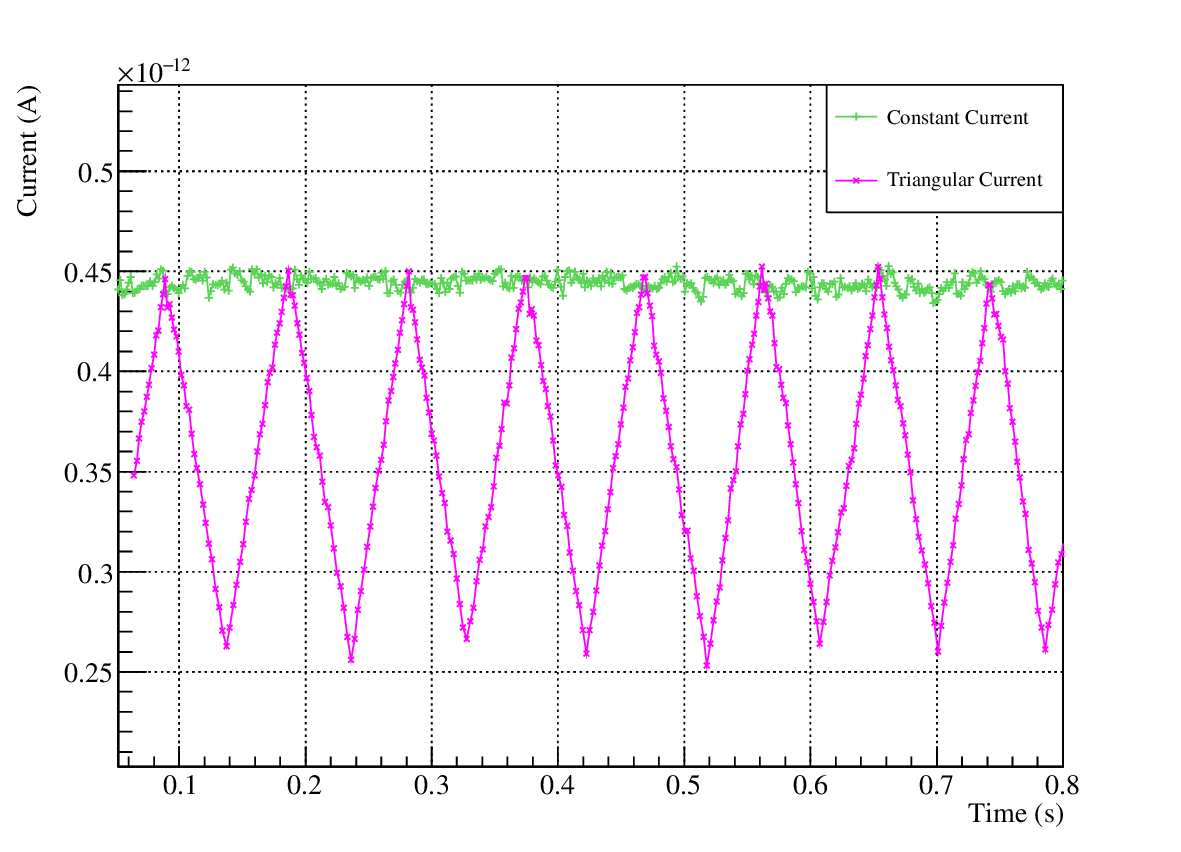}
        \caption{Output currents.}
        \label{fig:currents_CCD}
    \end{subfigure}
    \caption{Experiment to control the delay before draining the charge packets based on the charge value.}
    \label{fig:arbitrary_current}
\end{figure}

The first 500 pixels of the array were used for the experiment. The charge was produced by exposing the sensor to a LED light source. Figure \ref{fig:charge_delaycicles} shows on the left axis the distribution of charge per pixel (average of 10 experiments). To obtain a constant current each pixel must be read with a pixel time that is proportional to the charge of the pixel. In our setup, this time must be an integer multiple of $29$ $\mu$s. The first target was to obtain a constant current of $457.93$ fA given the charge distribution in \ref{fig:charge_delaycicles}.

The number of delay cycles per pixel was adjusted to get the desired current based on the measurements of the pixels when exposed to light. The green square markers in figure \ref{fig:charge_delaycicles} show the calculated number of cycles per pixel. At this point, the sensor is exposed again, and the pixels are read using the pre-loaded number of cycles per pixel in the readout system. Figure \ref{fig:currents_CCD} depicts green  the measured output current by the CCD, showing a good agreement with its expected value. 

Then, the experiment was repeated, but the number of cycles was adjusted to obtain a triangular current in the sensor output. The number of cycles per pixel is depicted in the pink curve using right y-axis in figure \ref{fig:charge_delaycicles}, and the resultant current as measured by the CCD is shown in the pink-colored curve of figure \ref{fig:currents_CCD}.

\section{Discussion}
\label{sec:discussion}

In Section \ref{comparisson_results} the random variables for the electrometer and Skipper-CCD, denoted as $X_{\mathrm{elec}}$ and $X_{\mathrm{CCD}}$ respectively, are expected to have two main components: one associated with the natural Poisson statistics of photon arrivals $X_{\mathrm{light}}$, and the other associated with measurement and processing errors $X_{\mathrm{meas+proc}}$. Consequently, the expression for the total variance for each device is $\sigma^2 = \sigma_{\mathrm{light}}^2 + \sigma_{\mathrm{meas+proc}}^2$. Insights into this are evident from the first row of the table \ref{tab:stat_results}, which shows results with no light exposure. As the readout noise of the Skipper-CCD for this experiment is $\approx 5.7$ $e^-$, which is equivalent to $9.13\times10^{-19}$ C, these measurements shows almost zero variance for zero exposure. Then, the uncertainty of the CCD turns out to be negligible, thus setting the $\sigma_{\mathrm{meas+proc}}\approx 0$ and $\sigma_{\mathrm{CCD}} \approx \sigma_{\mathrm{light}}$. On the other hand, the electrometer exhibits a higher standard deviation of $\sigma = 3.08$ pC, mainly due to $\sigma_{\mathrm{meas+proc}}$. In the $200$ nC range used for all measurements, the instrument has a resolution of $1$ pC and an accuracy of $0.4\%$ of the reading value plus $5$ pC. 

As the exposure time increases, the mean charge increases as well. The standard deviation for the Skipper-CCD also increases, following $\sigma_{\mathrm{light}}$, as expected with increasing light exposure. However, the standard deviation of the electrometer remains approximately constant, as it is dominated by $\sigma_{\mathrm{meas+proc}}$, except for the longer exposure of $30$ seconds for which both variances are comparable.

The standard deviation of the difference, $\sigma_{\mathrm{diff}}$ in table \ref{tab:stat_results}, is consistently equal or lower than that of the electrometer. If the random variables associated with the devices are correlated, then the standard deviation of the difference between the distributions is expected to be 
\setlength{\mathindent}{130pt} 
\begin{equation}
\sigma_{\mathrm{diff}} = \sqrt{\sigma_{\mathrm{elec}}^2 + \sigma_{\mathrm{CCD}}^2 - 2\mathrm{Cov}(X_{\mathrm{elec}},X_{\mathrm{CCD}})}. 
\label{eq:sigmaDiff1}
\end{equation}
Considering that the standard deviation for the sensor corresponds to that of the light ($\sigma_{\mathrm{CCD}}\approx \sigma_{\mathrm{light}}$), then
\setlength{\mathindent}{80pt}
\begin{eqnarray}
 2 \mathrm{Cov}(X_{\mathrm{elec}},X_{\mathrm{CCD}}) & \approx  2\mathrm{Cov}(X_{\mathrm{light}}+X_{\mathrm{meas+proc}},X_{\mathrm{light}})  \nonumber   \\
 & = 2\mathrm{Cov}(X_{\mathrm{light}},X_{\mathrm{light}}) +2\mathrm{Cov}(X_{\mathrm{meas+proc}},X_{\mathrm{light}}) \nonumber  \\
 & =2 \sigma_{\mathrm{light}}^2 = 2 \sigma_{\mathrm{CCD}}^2
 \label{Eq:2Cov}
\end{eqnarray}
where we used that $\mathrm{Cov}(X_{\mathrm{meas+proc}},X_{\mathrm{light}})=0$ due to independence. Replacing equation (\ref{Eq:2Cov}) in (\ref{eq:sigmaDiff1}) results in
\setlength{\mathindent}{160pt}
\begin{eqnarray}
	\setlength{\mathindent}{130pt}
   \sigma_{\mathrm{diff}} &\approx \sqrt{\sigma_{\mathrm{elec}}^2 - \sigma_{\mathrm{CCD}}^2}  \nonumber \\
\end{eqnarray}
which is compatible with the observed results as can be verified in table \ref{tab:stat_results}. This supports the assumption that the Skipper-CCD measurements are related only to the light measurements.

The mean of the difference, $\mu_{\mathrm{diff}}$, shows a small increasing difference with exposure. This discrepancy is likely due to a systematic error in the gain of the electrometer, which is not calibrated and/or its post-processing to remove the drift. A linear regression to quantify this systematic offset was performed between the mean of both instruments, yielding a coefficient of determination ($R^2$) very close to one ($0.9998$), confirming a strong linearity. In addition, the residuals have a mean of zero and are within $\pm 2s_{yx}$, where $s_{yx} = 0.74$ (standard deviation of residuals), also suggesting that the model fits the data adequately. The slope obtained for the linear regression, $m=1.026$, was used as a correction: $\mu^{*}_{\mathrm{diff}} = \mu_{\mathrm{elec}}/m - \mu_{\mathrm{CCD}}$,  and added as the last column in table \ref{tab:stat_results}. This corrected mean is compatible with the errors presented in the table.

The measurements presented in Section \ref{subsec:nano-Ampere} are dominated by Poisson statistics due to the nature of photon arrivals. The proportional increase in current is consistent with what is expected as the exposure increases. The exposure of 80 s shows that a current of $1$ nA is achievable by a single sensor, and measurements from the electrometer and the Skipper-CCD are compatible. Even larger currents could be achieved by faster readout, which requires optimization of the bandwidth of the readout electronic of the Skipper-CCD, by adjusting the sensor voltages to achieve a larger full-well capacity, or by using new Skipper sensors, such as those described in Section \ref{sec:ppm}.

The experiment in Section \ref{sec:arbitraryCurrentGeneration} was intended to demonstrate the flexibility of the output stages in adjusting the output current using the charge packet release time. In an actual application, the readout system should be able to change this timing as a function of the current and previously measured charge packets. Although, in the case presented, we calculate these time stamps using offline processing, online calculations have been implemented in the past to increase the sensor's sensitivity \cite{chierchie2020smartreadout}. At the same time, with the current readout hardware, time delays of $66$ ns are easily implementable and adjustments of $10$ ns are achievable with firmware modifications. A dedicated hardware implementation could get even more refined time shaping for future tests. 

\section{Conclusions} 

This paper presented the single-electron resolution Skipper-CCD as a potential candidate for implementing a current source aligned with the recent redefinition of the ampere. The charge measured by the Skipper-CCD was compared with that of an electrometer, and additional experiments were conducted to demonstrate the sensor's versatility and precision in managing charge packets and enabling self-calibration. Although slower than other technologies, the Skipper-CCD's capacity to handle electron packets across a large dynamic range, along with advancements in new CMOS-based Skipper sensors, featuring one amplifier per pixel, and multi-amplifier-sensing MAS-CCDs with Skipper capabilities in several in-line amplifiers, makes this technology a possible candidate for quantum current source generation.

\section*{Ackkowledgments}
Lawrence Berkeley National Laboratory (LBNL) is the developer of the fully-depleted CCD and the designer of the Skipper readout. The CCD development work was supported in part by the Director, Office of Science, of the U.S. Department of Energy under Contract No. DE-AC02-05CH11231.

The work of the research group at IIIE-UNS-CONICET was partially supported by projects PIBAA-0162 and PGI 24/K090, from CONICET and Universidad Nacional del Sur.

\section*{References}
\bibliographystyle{iopart-num}
\bibliography{main}

\end{document}